\documentclass[12pt]{article}%
\usepackage{amsmath}
\usepackage{amsfonts}
\usepackage{amssymb}
\usepackage{graphicx}%
\providecommand{\U}[1]{\protect \rule{.1in}{.1in}}
\allowdisplaybreaks

\usepackage[colorlinks=true, bookmarksopen=true,pdfstartview=FitH]{hyperref}
\hypersetup{bookmarksnumbered, colorlinks=true,linkcolor=blue,
citecolor=blue, urlcolor=blue}

\begin{document}

\title{{\Huge Non-uniform Black Strings with Schwarzschild-(Anti-)de Sitter
Foliation}\thanks{Work supported by the National Natural Science Foundation of
China through grant No. 90403014.}}
\author{Liu Zhao\thanks{E-mail: \texttt{lzhao@nankai.edu.cn}}
\hspace{5mm}
Kai Niu\hspace{5mm} Bing-Shu Xia\hspace{5mm} Yi-Ling Dou \hspace{5mm} Jie Ren
\\Department of Physics, Nankai University, 
\\Tianjin 300071, P R China}
\date{}
\maketitle

\begin{abstract}
We present some exact non-uniform black string solutions of 5-dimensional pure Einstein gravity as 
well as Einstein-Maxwell-dilaton theory at arbitrary dilaton coupling. The solutions share the
common property that their 4-dimensional slices are Schwarzschild-(anti-)de Sitter spacetimes.
The pure gravity solution is also generalized to spacetimes of dimensions higher than 5 to get 
non-uniform black branes.
\end{abstract}

\section{Introduction} 

Gravitational theories with extra spacetime dimensions have been a fascinating
area of study for a number of reasons. Besides perspectives from string
theory, Kaluza-Klein theory, braneworld scenario, and holography principle,
the non-uniqueness of black holes in higher dimensions and the existence of
solutions with non-spherical or even non-compact event horizons render the
higher dimensional gravitational theory an important branch of modern
relativity in its own right. More than ten years ago, Gregory and Laflamme
published their famous work \cite{Gregory-Laflamme} on the stability issues of black branes (strings).
They found that black branes with uniform stretched horizons in Ricci flat
spacetimes are generally unstable against classical perturbations of the
metric. This work leads to vast subsequent researches on related problems. One
of the corresponding developments is on the final fate of the black
branes/strings due to the classical perturbations. For instance, it has been
generally believed that black strings would segregate into an array of black
holes due to the Gregory-Laflamme instability. However, this picture has
changed drastically due to the innovative work of Horowitz and Maeda \cite{Horowitz-Maeda}, who
revealed that the the horizon of a black string cannot pinch off at finite
affine parameter and so the string would not segregate into black holes. In
order to have more detailed understandings of the fate of black strings, a 
considerable amount of works have been made, see e.g. 
\cite{Choptuik-etal, Tamaki-etal, 0601079, grqc0609046, 0702053, Hirayama-Kang}. 
For detailed review on the subject
and a much more complete list of references, see the review articles \cite{Kol} and \cite{HNO}. 
Most of the related works rely on numerical methods, because
the intermediate states of a uniform black string evolving toward its final
(stable) states are generally non-uniform, and it is extremely difficult to
study non-uniform black strings analytically. On the other hand, there indeed
exist some exact non-uniform black string solutions in the literature, and
some of these are even free of Gregory-Laflamme instabilities \cite{ grqc0609046, Hirayama-Kang}. 

In this article, we shall present a new exact non-uniform black string
solution in a 5-dimensional Ricci flat spacetime. The basic properties of the
solution as well as its behavior under small classical perturbations will be
analyzed. It turns out that the solution is unstable against classical
perturbations. Adding 3-branes into the solution makes the situation a little
bit better, but the lack of confining box in the fifth dimension implies that 
the instability might still persists.

\section{The vacuum solution}

\subsection{Deriving the solution}

The aim of the present section is to find some exact non-uniform black string solutions in
5D pure Einstein theory. For this we adopt the following metric ansatz%
\begin{align}
d\hat{s}^{2}  &  =g_{MN}(X)dX^{M}dX^{N}=e^{B(z)}\gamma_{\mu \nu}(x)dx^{\mu
}dx^{\nu}+e^{-C(z)}dz^{2},\label{ds}\\
\gamma_{\mu \nu}(x)dx^{\mu}dx^{\nu}  &  =-f(r)dt^{2}+\frac{dr^{2}}{f(r)}%
+r^{2}\left(  d\theta^{2}+\sin^{2}\theta d\varphi^{2}\right)  , \label{ds2}%
\end{align}
where coordinates for the full 5D spacetime are labeled by capital Latin
entices while those for the 4D subspaces at fixed fifth coordinate $z$ is
labeled by Greek indexes, so $X^{M}=(x^{\mu},z)$ in general. Direct
calculations show that the only non-vanishing components of the Ricci tensor
are%
\begin{align}
\hat{R}_{00}  &  =\frac{f(r)}{4r}\left[  2rf^{\prime \prime}(r)+4f^{\prime
}(r)+r\left(  2B^{\prime \prime}(z)+4B^{\prime}(z)^{2}+B^{\prime}(z)C^{\prime
}(z)\right)  e^{B(z)}e^{C(z)}\right]  ,\label{r00}\\
\hat{R}_{11}  &  =-\frac{1}{f(r)^{2}}\hat{R}_{00},\nonumber \\
\hat{R}_{22}  &  =-rf^{\prime}(r)+1-f(r)-\frac{1}{4}r^{2}\left(
2B^{\prime \prime}(z)+4B^{\prime}(z)^{2}+B^{\prime}(z)C^{\prime}(z)\right)
e^{B(z)}e^{C(z)},\label{r22}\\
\hat{R}_{33}  &  =\left(  \sin^{2}\theta \right)  \hat{R}_{22},\nonumber \\
\hat{R}_{44}  &  =-\left(  2B^{\prime \prime}(z)+B^{\prime}(z)^{2}+B^{\prime
}(z)C^{\prime}(z)\right)  , \label{r44}%
\end{align}
where we denote geometric quantities associated with the full spacetime metric
$g_{MN}$ by hatted symbols, while those associated with $\gamma_{\mu \nu}$ will
be un-hatted.

The form of the non-vanishing components for $\hat{R}_{MN}$ suggests that
we can get exact solution to the vacuum Einstein equation. For this we first
solve $\hat{R}_{44}=0$ for $C(z)$, getting%
\begin{align}
B^{\prime}(z)C^{\prime}(z)  &  =-2B^{\prime \prime}(z)-B^{\prime}%
(z)^{2},\label{bc}\\
C(z)  &  =-B(z)-2\log B^{\prime}(z)+\log \left(  \frac{4\Lambda}{3}\right)
,\quad if\quad B^{\prime}(z)\neq0, \label{c}%
\end{align}
where $\Lambda$ is provided as an integration constant. Inserting (\ref{bc})
and (\ref{c}) into (\ref{r00}) and (\ref{r22}), we find that both identities
are now independent of $z$, and a common solution to the equations $\hat
{R}_{00}=0$ and $\hat{R}_{22}=0$ can be easily found to be%
\begin{equation}
f(r)=1-\frac{2M}{r}-\frac{\Lambda r^{2}}{3}, \label{Cc}%
\end{equation}
where again, $M$ is to be interpreted as an integration constant. One
immediately recognize that, eq. (\ref{ds2}), equipped with (\ref{Cc}), is just
the metric of the well-known Schwarzschild-(A)dS black hole, with $M$
representing the Schwarzschild mass and $\Lambda$ being the cosmological
constant. Note that none of these constants are put in by hand -- they all
arise as integration constants.

Now any pair of functions $B(z)$ and $C(z)$ obeying (\ref{bc}) would make
(\ref{ds}) an exact solution of the 5D vacuum Einstein equation $\hat{R}%
_{MN}=0$. Since the 4D part is a black hole and the solution extends along the
fifth dimension $z$, the solution describes a non-uniform black string in 5D
Ricci flat spacetime.

Why there is only one constraint equation for the two functions $B(z)$ and
$C(z)$? The answer is clearly related to the coordinate choice freedom for the
coordinate $z$. For simplicity we may choose a coordinate in which
\begin{equation}
C(z)=-B(z) + iK\pi, \quad K = 0 \mbox{ or } 1. \label{K}
\end{equation}
The reason for introducing the constant $K$ 
will be clear instantly. Inserting (\ref{K}) into (\ref{bc}) and (\ref{c}), one gets the explicit form 
of the solution
\begin{align}
& d\hat{s}^{2}=e^{B(z)}\left[  -f(r)dt^{2}+\frac{dr^{2}}{f(r)}+r^{2}\left(
d\theta^{2}+\sin^{2}\theta d\varphi^{2}\right)  +dz^{2}\right]  , \nonumber\\
& f(r)=1-\frac{2M}{r}-\frac{\Lambda r^{2}}{3}, \quad B(z)=e^{-iK\pi/2} \sqrt{\frac{4\Lambda}{3}}z, \label{mt}%
\end{align}
where a further additive integration constant is omitted by a shift in the
coordinate $z$. It turns out that in order to keep the metric real, we need to take the constant 
$K=0$ for $\Lambda>0$ (de Sitter case) and $K=1$ for $\Lambda<0$ (anti-de Sitter case). In either cases 
we always have 
\begin{equation}
B(z)=\frac{2}{3}\sqrt{3|\Lambda|} z. \label{Bfunc}
\end{equation}
So the 4D subspace described by the metric $\gamma_{\mu \nu}$ is a
Schwarzschild-(A)dS spacetime, and the full metric 
$g_{MN}$ can be interpreted as a Ricci flat spacetime with Schwarzschild-(A)dS foliation. Note that 
from the 5D point of view $\Lambda$ does not have any special meaning: different values of $\Lambda$ 
just correspond to different ways of foliating the same 5D Ricci flat spacetime. 

The sturcture of the spacetime (\ref{mt}) is best described by the value of its curvature invariants. Direct
calculations yield
\begin{align*}
&\hat{R}^{MNPQ}(g)\hat{R}_{MNPQ}(g)=e^{-2B(z)}
\left[ R^{\sigma\mu\nu\rho}(\gamma)R_{\sigma\mu\nu\rho}(\gamma)- \frac{2|\Lambda|}{3}R(\gamma) \right]\\
&\qquad=\frac{8}{3}\exp\left(-\frac{4}{3} \sqrt{3|\Lambda|}z\right)
\left[ \frac{18M^{2}}{r^{6}} + (\Lambda -|\Lambda|)^{2} \right].
\end{align*}
The center of the black string, i.e. $r=0$ is a usual singularity which is surrounded by the event 
horizons at the roots of $f(r)$. In addition to this, at $z \rightarrow -\infty$, there is 
also a singularity which is out side of any event horizon. The same kind of naked singularities also 
appeared in the study of braneworld black holes \cite{9909205} and in the uniform Schwarzschild-dS black 
string in 5-dimensions \cite{Hirayama-Kang}, and in the latter case Gregory \cite{Gregory} has pointed 
out that the existence of such a singularity is closely related to the unstability of such black strings. 
In the next subsection we shall make similar discussions to \cite{Gregory} for the Ricci flat black string
(\ref{mt}) we just obtained above.

It should be reminded that the above solution corresponds only to the case
$B^{\prime}(z)\neq0$. For the case in which $B^{\prime}(z)=0$, it is not
unreasonable to set $B=0$. Then $C(z)$ can be any analytic function of $z$ and
the full solution corresponds to a uniform black string provided $\gamma
_{\mu \nu}(x)$ is given by any 4D Ricci flat black hole solution. This latter
case will be discarded in the rest of this article since it is a well studied
subject in the literature.

It should also be reminded that the coordinate condition (\ref{K}) is not
the only choice and not even the most interesting choice. We can choose, for
instance, another coordinate $\zeta$ for the fifth dimension, with
$d\zeta=e^{B(z)/2}dz$. Under such a coordinate, the metric becomes\footnote{
We noted while finishing this article that this form of the metric has already been known
and used to discuss the necesity for the existence of a 4D cosmological constant
from a 5D Kaluza-Klein perspective \cite{Wesson, Romero}. The same form of the solution 
can also be recovered as special $k=1$ case of eqs.(6.55), (6.66) in \cite{0603177}. 
Our point of view on the 5D solution is very different from those works. We thank 
R. Troncoso for communication on the latter point. } 
\[
d\hat{s}^{2}=\frac{|\Lambda|}{3}\zeta^{2}\left[  -f(r)dt^{2}+\frac{dr^{2}}%
{f(r)}+r^{2}\left(  d\theta^{2}+\sin^{2}\theta d\varphi^{2}\right)  \right]
dx^{\mu}dx^{\nu}+d\zeta^{2}.
\]
It is then interesting to see that, if we set $M=0$ and $\Lambda=3k, k=\pm 1$, rename
$t$ into $y$, and make the following double Wick rotation%
\[
\zeta\rightarrow i\tau,r\rightarrow ir,
\]
the metric will become%
\[
d\hat{s}^{2}=-d\tau^{2}+\tau^{2}\left[  \left(  1-k r^{2}\right)  ^{-1}%
dr^{2}+r^{2}\left(  d\theta^{2}+\sin^{2}\theta d\varphi^{2}\right)  +\left(
1-k r^{2}\right)  dy^{2}\right]  .
\]
It turns out that the constant $y$ slices of this metric are 4D FRW metrics
with spacial curvature $k$. Such a metric can be obtained as solutions of
the standard ideal gas cosmology with equation of state parameter
$\omega=-2/3$. From the point of view of Kaluza-Klein theory, the $yy$ component
of the metric plays the role of a scalar in the 4-dimensional subspace spanned
by the coordinates $(\tau, r, \theta, \phi)$, so the acceleration in the 4-dimensional
subspace is triggered by a \emph{spatially-varying} rather than temporally-varying scalar field.
Since our main goal is on the black string interpretation of
the solution, we shall not expand further on the double Wick rotated form of
the metric.

\subsection{Gregory-Laflamme instability of the vacuum solution} \label{GLa}

Now we return to the metric (\ref{mt}) and study its behavior under small
perturbations. For uniform black string in Ricci flat spacetimes and
non-uniform black strings in 5D AdS spacetimes, similar analysis were made
respectively in \cite{Gregory-Laflamme} and \cite{Hirayama-Kang} and the results turns out to be very
different. So it is interesting to see the behavior of our solution under
similar perturbations.

The idea is to perturb the metric in such a way that%
\begin{align}
g_{MN}  &  \rightarrow g_{MN}+\delta g_{MN},\nonumber \\
d\hat{s}^{2}  &  \rightarrow d\hat{s}^{\prime2}=e^{B(z)}\left(  \left(
\gamma_{\mu \nu}+e^{-B(z)}h_{\mu \nu}(x,z)\right)  dx^{\mu}dx^{\nu}%
+dz^{2}\right)  , \label{pert}%
\end{align}
i.e.%
\begin{align*}
\delta g_{MN}  &  =h_{MN}(x,z),\\
h_{4\mu}  &  =h_{44}=0.
\end{align*}
Then up to terms linear in $h_{\mu \nu}(x,z)$, the perturbed Einstein equation
reads (Palatini identity)%
\begin{equation}
\delta R_{MN}(g)=\frac{1}{2}g^{KL}\left(  \hat{\nabla}_{N}\hat{\nabla}%
_{M}h_{KL}-\hat{\nabla}_{L}\hat{\nabla}_{N}h_{KM}-\hat{\nabla}_{L}\hat{\nabla
}_{M}h_{KN}+\hat{\nabla}_{L}\hat{\nabla}_{K}h_{MN}\right)  =0, \label{linear}%
\end{equation}
where $\hat{\nabla}_{N}$ denotes the covariant derivative associated with the
unperturbed 5D metric $g_{MN}$. In order to solve the perturbation equation,
we have to supplement the equation (\ref{linear}) with proper gauge
conditions. As usual, the gauge conditions can be chosen to be transverse
traceless and Lorentzian, i.e.
\begin{align*}
h  &  \equiv \gamma^{\mu \nu}h_{\mu \nu}=0,\\
\nabla_{\mu}h^{\mu \nu}  &  =0,
\end{align*}
where $\nabla_{\mu}$ represent covariant derivatives associated with
$\gamma_{\mu \nu}$. Imposing the above conditions and after long, tedious and
somewhat standard calculations, the perturbation equation can be rearranged
into the following form,%
\begin{equation}
\square^{(\gamma)}h_{\mu \nu}+2R_{\mu \rho \nu \lambda}(\gamma)h^{\rho \lambda
}+\left[  \partial_{z}^{2}-\frac{1}{2}B^{\prime}(z)\partial_{z}-B^{\prime
\prime}(z)-\frac{1}{2}B^{\prime}(z)^{2}\right]  h_{\mu \nu}=0, \label{okp}%
\end{equation}
where $\square^{(\gamma)}$ is the D'alembertian and $R_{\mu \rho \nu \lambda
}(\gamma)$ is the Riemann tensor associated with $\gamma_{\mu \nu}$.

To further analyze the equation (\ref{okp}), we make the following separation
of variables,%
\begin{equation}
h_{\mu \nu}(x,z)=h_{\mu \nu}(x)e^{B(z)/4}\xi(z). \label{sep}%
\end{equation}
Then (\ref{okp}) becomes%
\begin{align}
\left[  \square^{(\gamma)}h_{\mu \nu}(x)+2R_{\mu \rho \nu \lambda}(\gamma
)h^{\rho \lambda}(x)\right]   &  =m^{2}h_{\mu \nu}(x),\label{ho}\\
\left[  -\partial_{z}^{2}+\frac{3}{4}B^{\prime \prime}(z)+\frac{9}{16}%
B^{\prime}(z)^{2}\right]  \xi(z)  &  =m^{2}\xi(z), \label{sch}%
\end{align}
where $m^{2}$ is a constant arising from the separation of variables.
Inserting $B(z)=2\sqrt{\frac{|\Lambda|}{3}}z$ (eq. \ref{Bfunc})) into (\ref{sch}), we get
\begin{equation}
\left[  -\partial_{z}^{2}+\frac{3|\Lambda|}{4}\right]  \xi(z)=m^{2}\xi(z), \label{5th}
\end{equation}
which is a Schr\"odinger equation with a constant potential $V(z)=\frac
{3|\Lambda|}{4}$. The normalizable solutions to this equation are plane waves
with continuous spectrum, with eigenvalues $m^{2}>\frac{3|\Lambda|}{4}$. From
the effective 4D point of view, the physical interpretation of the constant
$m$ is, in the spirit of Kaluza-Klein theory, the mass of the Kaluza-Klein
gravitons. So we are left with eq. (\ref{ho}) with $m^{2}$ having a non-zero
minimum. This is the famous massive Lichnerowitz equation. Under Schwarzschild-(A)dS
background, the solution of this equation was studied by Hirayama and Kang 
in \cite{Hirayama-Kang}. They showed that the Schwarzschild-de Sitter case is always 
unstable while the Schwarzschild-AdS case can be stable provided the length of the 
string exceeds the AdS radius. In our case the perturbation equation (\ref{5th}) in the fifth 
dimension is different from the cases studied by Hirayama and Kang. Rather, it is 
almost identical to the case studied in Gregory and Laflamme's original paper, the only 
difference being a non-zero minimum of the constant potential. So the behavior of our
solution under small perturbations of the form (\ref{pert}) is basically the
same as the uniform black string and is very different from the cases of AdS
black string studied by Hirayama and Kang. In the latter case, the effective
Schr\"odinger equation arising from separation of variables of the perturbation
equations contains an effective confining potential which prevents the
perturbation from evolving far away and thus the black string can be stable to
some extent. In our case, the effective potential is a constant, and any
perturbation will propagate along the $z$ axis infinitely. The only effect of
the constant potential is to make the Kaluza-Klein mass $m$ having a non-zero
minimum, and this is exactly the requirement of a physical perturbation (i.e.
any perturbation of the form (\ref{pert}) is physical in our case).

\begin{figure}[tbh]
\begin{center}
\includegraphics[width=2.5123in]{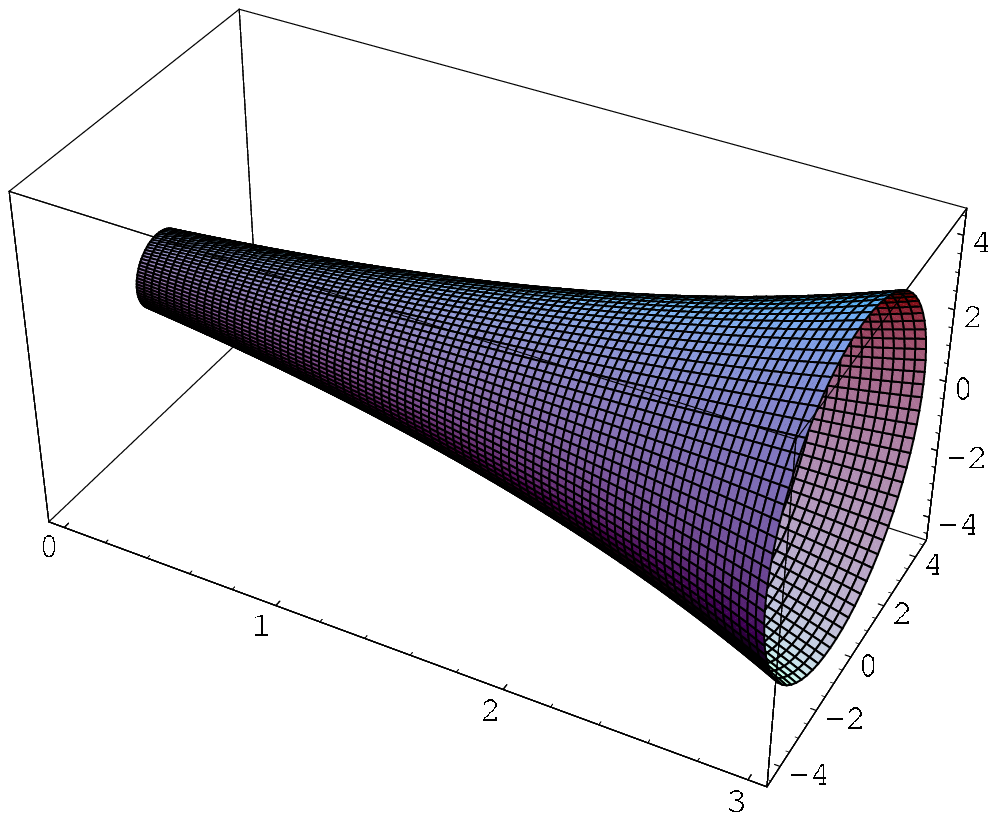}
\includegraphics[width=2.5123in]{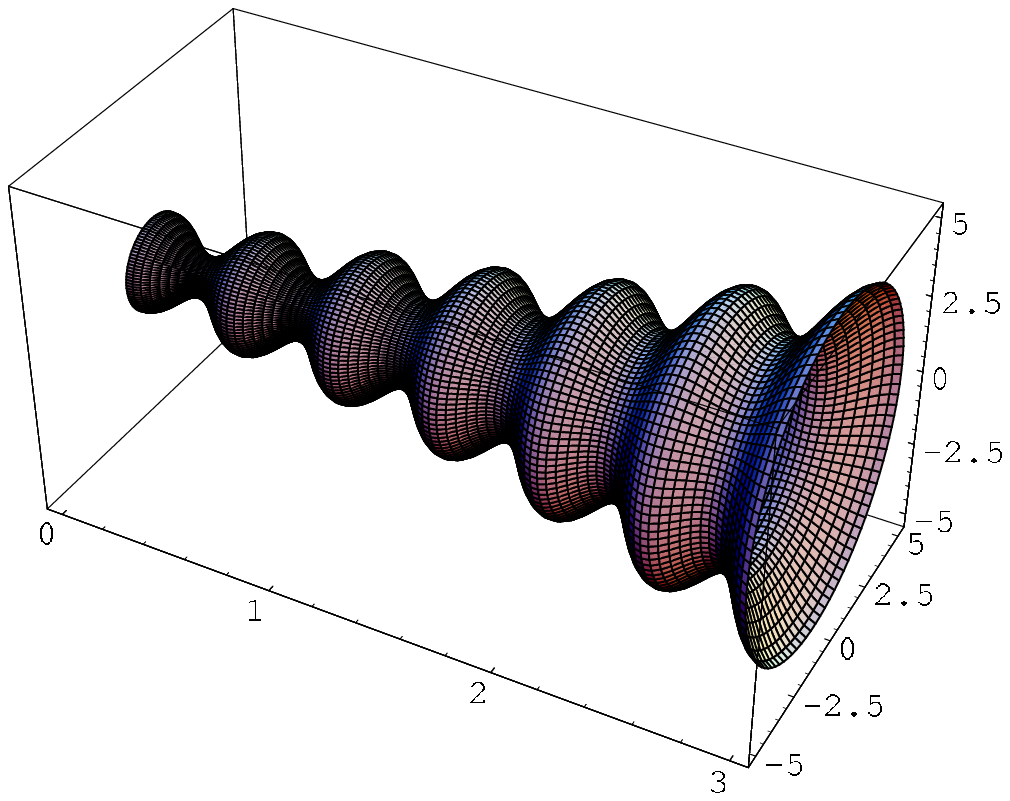}
\end{center}
\caption{Horizon of the black string before and after a single $s$-wave perturbation}\label{fig1}
\end{figure}

It might be illustrative to give a pictorial presentation of the horizons of
our black string solution both before and after the perturbation. Figure
\ref{fig1} is presented for this purpose. Note that the perturbation shown in
Figure \ref{fig1} is only a single mode perturbation. Generic perturbations are an
arbitrary superposition of all such modes with different, continuously
distributed wave lengths. So the perturbed horizon can look quite irregular
along the $z$ axis. However the constant $z$ slices are all Schwarzschild-(A)dS black holes,
since we are considering only spherical symmetric perturbations.

%\begin{figure}[tbh]
%\includegraphics[height=90mm]{unperturbed.eps}
%\includegraphics[height=90mm]{perturbed.eps} \centerline{(a)\hspace{6cm}(b)}
%\caption{ The horizons of the new black string solution, before (a) and after
%(b) perturbation}%
%\label{fig1}%
%\end{figure}%

Before ending this subsection we address that even in the Schwarzschild-AdS case where a stable threshold mass
exists as shown by Hirayama and Kang \cite{Hirayama-Kang}, the solution described in  this section cannot 
reach a new static configuration. This is because that although at the threshold mass the 
time-damping/increasing modes are 
abscent, the oscillating modes, however, can travel in the fifth dimension without barrier due to the flat
potential appearing in (\ref{5th}). Therefore we do not expect new static configurations to emerge after the 
perturbation.

\subsection{Adding Brane}

It is nowadays a common practice to consider black strings in higher
dimensions attached to some brane configurations. Normally adding such
brane(s) will introduce extra attractive potentials around the brane. So we
would like to see whether adding a brane to our black string solutions will
make the string more stable.

The key point in introducing a single brane into the existing solution of
Einstein equation is to make the solution $Z_{2}$-symmetric around certain
point along the fifth coordinate axis. In our case, we could simply change the
function $B(z)$ in (\ref{mt}) into its absolute value, i.e.
\begin{equation}
d\hat{s}^{2}=e^{|B(z)|}\left[  -f(r)dt^{2}+\frac{dr^{2}}{f(r)}+r^{2}\left(
d\theta^{2}+\sin^{2}\theta d\varphi^{2}\right)  +dz^{2}\right] , \label{ma}%
\end{equation}
where $B(z)$ is still given by eq.(\ref{Bfunc}). It is straightforward to show 
that the above metric is an exact solution to
the equation%
\begin{equation}
M_{\ast}^{3}\sqrt{g}\left[  \hat{R}_{MN}-\frac{1}{2}g_{MN}\hat{R}\right]
=T\sqrt{\gamma}\delta_{M}^{\mu}\delta_{N}^{\nu}\gamma_{\mu \nu}\delta(z),
\label{beq}%
\end{equation}
which is the equation of motion arising from variation of the action%
\[
S=-\frac{M_{\ast}^{3}}{2}\int d^{4}x\int dz\sqrt{g}\hat{R}-T\int d^{4}%
x\sqrt{\gamma}.
\]
Here $T$ is to be interpreted as the tension of a $3$-brane located at $z=0$,
and in order that (\ref{ma}) solves (\ref{beq}), $T$ must not be taken
arbitrarily but rather has the value%
\begin{equation}
T=2M_{\ast}^{3}\Lambda. \label{tension}%
\end{equation}

At first sight, this whole picture is very similar to the well-known RS2
braneworld model of Randall and Sundrum \cite{RS2}, however there are some fundamental
differences. The most important difference lies in that the RS2 model has an
AdS$_{5}$ bulk, while the present model has a 5D Ricci flat bulk.
Consequently, in RS2 the brane tension $T$ is fine-tuned with the bulk
cosmological constant, while the brane tension in our case is related to the
cosmological constant on the brane via (\ref{tension}), which is not a
fine-tuning condition since $\Lambda$ is only an integration constant before
the introduction of the brane and adding the brane simply fix this constant.

What we would like to see is whether the addition of the brane at $z=0$ makes
the solution more stable. For this purpose we need to repeat the procedure of
perturbation carried out in the last section, but this time the equation to be
perturbed becomes (\ref{beq}). Notice that the right hand side of (\ref{beq})
is non-vanishing only at $z=0$, and at the non-vanishing point, we have
$\gamma_{\mu \nu}=g_{\mu \nu}$. Therefore, we can rewrite (\ref{beq}) as%
\begin{equation}
\hat{R}_{MN}-\frac{1}{2}g_{MN}\hat{R}=2\Lambda \delta_{M}^{\mu}\delta_{N}^{\nu
}g_{\mu \nu}\delta(z), \label{eq1}%
\end{equation}
where the relation (\ref{tension}) between the brane tension and the 4d
cosmological constant has been made use of. Taking the trace of (\ref{eq1}),
we get%
\begin{equation}
-\frac{3}{2}\hat{R}=8\Lambda \delta(z), \label{lam}%
\end{equation}
which can be used to eliminate $\delta(z)$ on the right hand side of
(\ref{eq1}), yielding%
\begin{align}
\hat{R}_{\mu \nu}-\frac{1}{8}g_{\mu \nu}\hat{R}  &  =0,\label{al}\\
\hat{R}_{zz}-\frac{1}{2}g_{zz}\hat{R}  &  =0. \label{bl}%
\end{align}
We can then use (\ref{bl}) to solve $\hat{R}$ in terms of $\hat{R}_{zz}$ and
$g_{zz}$. Inserting the result into (\ref{al}), we get finally the following
equation,%
\begin{equation}
\hat{R}_{\mu \nu}-\frac{1}{4}g_{\mu \nu}(g_{zz})^{-1}\hat{R}_{zz}=0. \label{d}%
\end{equation}
So, the stability analysis in the presence of the brane located at $z=0$ would
be concentrated in the study of the behavior of (\ref{d}) under perturbations
of the form (\ref{pert}).

The perturbation equation can be written as%
\[
\delta \hat{R}_{\mu \nu}(g)-\frac{1}{4}h_{\mu \nu}(g_{zz})^{-1}\hat{R}_{zz}%
-\frac{1}{4}g_{\mu \nu}(g_{zz})^{-1}\delta \hat{R}_{zz}=0.
\]
After long and tedious calculations, the perturbation equation can be
rearranged into the following form,%
\begin{align*}
&  \square^{(\gamma)}h_{\mu \nu}+2R_{\mu \rho \nu \lambda}(\gamma)h^{\rho \lambda
}+\left[  \partial_{z}^{2}-\frac{1}{2}a^{\prime}(z)\partial_{z}-a^{\prime
\prime}(z)-\frac{1}{2}a^{\prime}(z)^{2}\right]  h_{\mu \nu}\\
&  \quad-\gamma^{\rho \sigma}\left[  R_{\rho \nu}(\gamma)h_{\sigma \mu}%
+R_{\sigma \mu}(\gamma)h_{\rho \nu}\right] +\frac{3}{2}\left[
a^{\prime \prime}+\left(  a^{\prime}\right)  ^{2}\right]  h_{\mu \nu}=0.
\end{align*}
where $a(z)\equiv|B(z)|$ and the second line is newly added comparing to the
case without the brane. As in the previous case we can make the separation of
variables%
\[
h_{\mu \nu}(x,z)=h_{\mu \nu}(x)e^{a(z)/4}\xi(z).
\]
Then the single perturbation equation becomes
\begin{align}
& \left[  \square^{(\gamma)}h_{\mu \nu}(x)+2R_{\mu \rho \nu \lambda}(\gamma
)h^{\rho \lambda}(x)\right]  \nonumber\\
& \qquad -\gamma^{\rho \sigma}\left[  R_{\rho \nu}%
(\gamma)h_{\sigma \mu}(x)+R_{\sigma \mu}(\gamma)h_{\rho \nu}(x)\right] 
=m^{2}h_{\mu \nu}(x), \label{changed}\\
& \left[  -\partial_{z}^{2}+\frac{3}{4}a^{\prime \prime}(z)+\frac{9}{16}%
a^{\prime}(z)^{2}\right]  \xi(z)-\frac{3}{2}e^{-a(z)/4}\left[  a^{\prime
\prime}(z)+a^{\prime}(z)^{2}\right] \xi(z) =m^{2}\xi(z).
\end{align}
Inserting $a(z)=\left|\sqrt{\frac{4\Lambda}{3}}z\right|$ into the second equation, we get%
\[
\left[  -\partial_{z}^{2}+V(z)\right]  \xi(z)=m^{2}\xi(z),
\]
where%
\begin{align}
V(z) &  =V_{a}(z)+V_{b}(z), \label{Vz}\\
V_{a}(z) &  =-|\sqrt{3\Lambda}|\delta(z), \label{Vz2}\\
V_{b}(z) &  =|\Lambda| \left[  \frac{3}{4}-2\exp \left(  -\frac{1}{2}\left|\sqrt 
{\frac{\Lambda}{3}}z \right|\right)  \right]  \mathrm{sign}^{2}(z).\label{Vz3}
\end{align}

\begin{figure}[tbh]
\begin{center}
\includegraphics[height=60mm]{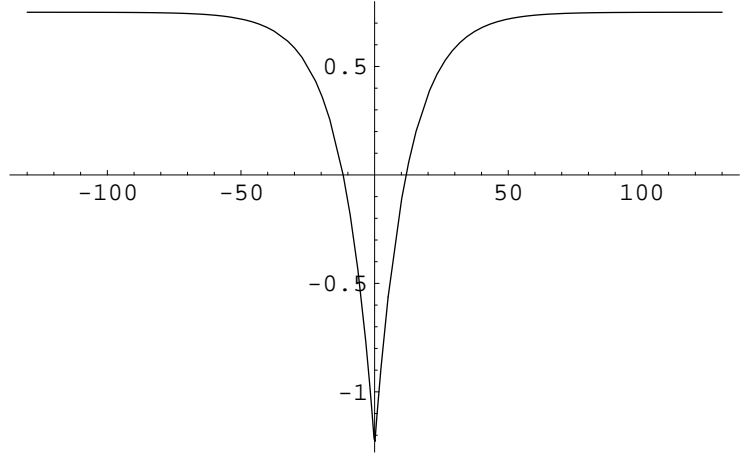}
\caption{ The effective potential around the brane}%
\label{fig2}%
\end{center}
\end{figure}%
Figure \ref{fig2} is devoted to the illustration of the effective potential $V(z)$. It can
be seen that there is an attractive potential well of
infinite depth around the brane at $z=0$, but far away the brane the potential approaches a
constant value $\frac{3|\Lambda|}{4}$. So the spectrum of the perturbation can
be divided into two different cases: for $m^{2}<\frac{3|\Lambda|}{4}$ there are
bound states around the brane, and hence these modes cannot propagate far
away; for $m^{2}>\frac{3|\Lambda|}{4}$, the spectrum is continuous and
unbounded, so any such mode is still unstable. So we have the conclusion that
though the presence of the brane effectively brings in an attractive potential
around the brane, the whole system can still be instable in general, depending on the nature of solution
to the first equation, eq.(\ref{changed}). Comparing to the standard Lichnerowitz equation (\ref{ho}),
eq.(\ref{changed}) has a different dispersion relation due to the appearance of the two extra terms. 
Thus in order to make final desision on the stability of the case with brane added, one needs to 
solve this new equation numerically. However, to us this is a very complicated story and we would 
just like to infer that the situation is very much like the case of braneword black holes \cite{9909205} 
except that we have now a Ricci flat bulk instead of an AdS one. In the literatures such as 
\cite{Hirayama-Kang, 9909205, Karch-Randall} the AdS bulk acting as a confining box is a strong evidence 
for the existence of stabe modes. The present case is on the contrary because we don't have such a confining box.
So most probably the addition of a brane in the Ricci flat bulk would not make the system stable. In 
particular, the pertubed system will never reach a new static form provided the strength of the 
perturbation exceeds the asymptotic value of $V(z)$ in (\ref{Vz})-(\ref{Vz3}) due to the same reason as described in the end of the last subsection.

\section{Solutions to Einstein-Maxwell-Dilaton theory}

The vacuum black string solution obtained earlier can be used to generate
black string solutions to Einstein-Maxwell-Dilaton theory. The explicit steps
are described as follows.

First we extend the 5D vacuum solution to 6D by simply adding a flat 6-th
dimension, i.e.
\[
ds_{6}^{2}=ds^{2}+dy^{2}.
\]
Then we apply a Lorentz boost in the $(t,y)$ plane,%
\begin{align*}
t  &  \rightarrow T=t\cosh \psi-y\sinh \psi,\\
y  &  \rightarrow Y=-t\sinh \psi+y\cosh \psi.
\end{align*}
After this step, the 6D metric becomes%
\begin{align*}
\hat{g}_{00}  &  =\cosh^{2}\psi \left[  1-e^{B(z)}f(r)\right]  -1,\\
\hat{g}_{05}  &  =\sinh \psi \cosh \psi \left[  1-e^{B(z)}f(r)\right]  ,\\
\hat{g}_{55}  &  =\cosh^{2}\psi \left[  1-e^{B(z)}f(r)\right]  +e^{B(z)}f(r),
\end{align*}
with the rest components unchanged.

Next, we make a Kaluza-Klein reduction along the 6-th coordinate dimension.
Recall that general K-K reduction formula is
\begin{equation}
ds_{D+1}^{2}=e^{2\alpha \phi}ds_{D}^{2}+e^{2\beta \phi}(dz+A)^{2}, \label{1}%
\end{equation}
i.e.%
\begin{align}
\hat{g}_{\mu \nu}  &  =e^{2\alpha \phi}g_{\mu \nu}+e^{2\beta \phi}A_{\mu}A_{\nu
},\nonumber \\
\hat{g}_{\mu z}  &  =e^{2\beta \phi}A_{\mu},\label{3}\\
\hat{g}_{zz}  &  =e^{2\beta \phi},\nonumber
\end{align}
where the parameters $\alpha,\beta$ are strictly restricted by the spacetime dimension under consideration:
\[
\alpha=\frac{1}{\sqrt{2(D-1)(D-2)}},\qquad \beta=-(D-2)\alpha=-\sqrt{\frac
{D-2}{2(D-1)}}.
\]
The corresponding Lagrangian for pure gravity reduces as%
\begin{equation}
\mathcal{L}=\sqrt{-\hat{g}}R_{(D+1)}=\sqrt{-g}\left(  R_{(D)}-\frac{1}%
{2}\left(  \partial \phi \right)  ^{2}-\frac{1}{4}e^{-\tilde\alpha \phi}%
F^{2}\right)  . \label{2}%
\end{equation}

At $D=5$ we have%
\[
\alpha=\frac{\sqrt{6}}{12},\beta=-3\alpha=-\frac{\sqrt{6}}{4}.
\]
For convenience, we also define%
\[
\tilde{\alpha}=2(D-1)\alpha=\sqrt{\frac{2(D-1)}{D-2}},
\]
which at $D=5$ takes the value%
\[
\tilde{\alpha}=\sqrt{\frac{8}{3}}=-\frac{8\beta}{3}.
\]

After the KK reduction, we get a solution to Einstein-Maxwell-dilaton theory at KK
coupling $\tilde\alpha=\sqrt{8/3}$. The explicit solution is given as%
\begin{align*}
g_{00}  &  =-V(r,z)^{-2/3}e^{B(z)}f(r),\\
g_{11}  &  =V(r,z)^{1/3}\frac{e^{B(z)}}{f(r)},\\
g_{22}  &  =V(r,z)^{1/3}e^{B(z)}r^{2},\\
g_{33}  &  =V(r,z)^{1/3}e^{B(z)}r^{2}\sin \theta,\\
g_{44}  &  =V(r,z)^{1/3}e^{B(z)},
\end{align*}
where%
\[
V(r,z)=\frac{e^{B(z)}f(r)k^{2}-1}{k^{2}-1},
\]
and
\[
k\equiv \tanh \psi
\]
represents the velocity of the boost mentioned above. The corresponding dilaton field and electric field potential are given
respectively by%
\begin{align*}
e^{2\beta \phi}  &  =V(r,z),\\
A  &  =V(r,z)^{-1}\frac{k\left[  e^{B(z)}f(r)-1\right]  }{k^{2}-1}dt.
\end{align*}
One often writes the dilaton solution in terms of the parameter $\tilde
{\alpha}$. In our case, we have%
\[
e^{-\tilde{\alpha}\phi}=\left(  e^{2\beta \phi}\right)  ^{4/3}=V(r,z)^{4/3}.
\]
In compact form, the solution is written as%
\[
d\tilde{s}^{2}=V(r,z)^{1/3}e^{B(z)}\left(  -V(r,z)^{-1}f(r)dt^{2}+\frac
{dr^{2}}{f(r)}+r^{2}d\theta^{2}+r^{2}\sin^{2} \theta d\varphi^{2}+dz^{2}\right)  .
\]

The solution obtained above corresponds to the Kaluza-Klein 
coupling $\tilde\alpha=\sqrt{8/3}$. To generalize the above solution to the case with 
arbitrary dilaton coupling $\tilde\alpha$, we introduce the following notation:%
\[
N=4/(\tilde{\alpha}^{2}+4/3).
\]
%We postulate (cf hep-th/0412153) that 
Then by straightforward calculations it can be checked that the following is a solution to the 
Einstein-Maxwell-dilaton theory at arbitrary dilaton coupling:%
\begin{align}
d\tilde{s}^{2} &  =V(r,z)^{N/3}e^{B(z)}\nonumber\\
&\quad\left(  -V(r,z)^{-N}f(r)dt^{2}%
+\frac{dr^{2}}{f(r)}+r^{2}d\theta^{2}+r^{2}\sin^{2} \theta d\varphi^{2}%
+dz^{2}\right)  , \label{finsol1}\\
A &  =\sqrt{N}V(r,z)^{-1}\frac{k\left[  e^{B(z)}f(r)-1\right]  }{k^{2}-1}dt,\label{finsol2}\\
e^{-\phi} &  =V(r,z)^{N\tilde{\alpha}/2}. \label{finsol3}
\end{align}
The corresponding Lagrangian is still represented by (\ref{2}) but with $\tilde\alpha$ arbitrary, 
with the equations of motion
\begin{align}
& R^{MN}-\frac{1}{2}\partial^{M}\phi \partial^{N}\phi-\frac{1}{2}e^{-\tilde
{\alpha}\phi}F^{MA}F_{A}{}^{N}+\frac{1}{4(D-2)}e^{-\tilde{\alpha}\phi}%
F^{2}g^{MN}=0,\\
& \square \phi+\frac{\tilde{\alpha}}{4}e^{-\tilde{\alpha}\phi}F^{2}=0,\\
& \left(  e^{-\tilde{\alpha}\phi}F^{MN}\right)  _{;M}=0
\end{align}
for $D=5$. 

It is tempting to take time to study some of the characteristic properties of the solution given 
in this section. First of all, it can be easily seen that at $k=0$, one has $V(r,z)=1$ and $A=0$, thus the 
whole system of solution reduces back to the original vacuum solution (\ref{mt}). Since the EMD theory 
at arbitrary dilaton coupling differ from the vacuum Einstein gravity by the presence of electromagnetic
field $A$ and the dilaton field $\phi$, one is naturally led to the understanding that the parameter $k$ 
must somehow be related to the electromagnetic charge of the solution. Indeed, one can see this by 
studying the Columb potential on the event horizon. Substituting the horizon condition $f(r)=0$ into 
the expressions for $A$ and $\phi$, we get
\begin{align*}
A &= \sqrt{N} k dt |_{f(r)=0},\\
e^{-\phi} &= \left(\frac{1}{1-k^{2}}\right)^{N\tilde\alpha/2}.
\end{align*}
We thus see that on the event horizon, not only the Coloumb potential remains a constant, but also the 
dilaton field is a constant. That the Coloumb potential is a constant on the horizons is a remarkable
fact, because usually there exist more than one horizons for the Schwarzschild-dS metric, and 
the present result implies that all these disjoint horizons have equal electric potential. Therefore, 
the electronic charge in this spacetime must be distributed in an unusual way. To derive the exact 
distribution of electronic charges however must involve global analysis of the spacetime, which 
is out of the scope of the present work.

Another aspect of the spacetime which is also of interest is the calculation of the surface gravity or 
Hawking temperature. There are quite a few different approaches in deriving the surface gravity of black holes, 
such as using the classical equation satisfied by the null Killing vector normal to the horizon(s), the 
Euclideanization technique or the tunneling effect through the hirizons etc. With the use of any of these 
methods it can be seen that the temperature of the solution (\ref{finsol1}-\ref{finsol3}) is
\[
T=\frac{1}{2\pi} \left(  1-k^{2}\right)  ^{N/6}\left \vert \frac
{M}{r_{H}^{2}}-\frac{\Lambda r_{H}}{3}\right \vert,
\]
where $r_{H}$ is the radius of the horizon on the $4$-dimensional Schwartzschild-(A)dS slices of 
the black string, i.e. the root of $f(r)$. For the Schwartzschild-dS branch there are 
more than one horizons, and the above formula gives the temperature for each horizon if appropriate 
horizon radius $r_{H}$ is inserted. Note that for each of the horizons the temperature is a constant
despide the fact that the horizons are themselves nonuniform.

The entropy of the black string which is one forth of the area of the horizon can also be calculated 
with ease. For this we have to mention that the solution (\ref{finsol1}-\ref{finsol3}) does not restrict 
the length of the string, so what we can actually calculate is only the entropy of a particular segment 
of the black string, extending in the $z$-direction from, say, $z_{0}$ to $z_{1}$. The entropy 
contained in this segment of the horizon is
\[
S=\pi r_{H}^{2} \left(  3\Lambda \right)^{-1/2} \left(  1-k^{2}\right)^{-N/2}
\left(e^{\sqrt{3\Lambda}z_{1}}-e^{\sqrt{3\Lambda}z_{0}}\right).
\]
Notice that this area is not proportional to the ``length'' $z_{1}-z_{0}$ of the string segment but 
is proportional to the particular expression $\left(e^{\sqrt{3\Lambda}z_{1}}-e^{\sqrt{3\Lambda}z_{0}}\right)$.
This is because that the black string we consider here is not translationaly invariant along $z$-direction,
thus the length of the string segment is not a good parameter characterizing the area of the horizon. Actually
it is this observation that prevented us from comparing the analysis of Gregory-Laflamme instability 
made in Subsection \ref{GLa} with the famous Gubser-Mitra conjecture \cite{GM}. The latter says that 
for black strings with translationary invariance Gregory-Laflamme instability appears if and only if 
it is thermodynamically unstable. It is interesting to ask whether the Gubser-Mitra conjecture still holds 
if the translationary invariant condition is removed.

\section{Black branes}
The vacuum solution described in section 2 can also be generalized  along a different line, i.e. 
to the case with more than one longitudinal dimensions.

For example, we can easily get a 6D vacuum black brane solution by starting from the metric ansatz
\[
ds^2 = B(y,z)\left( - f(r) dt^2 + \frac{dr^2}{f(r)} + r^2 d\Omega_2^2 \right) + C_1(y) dy^2 + C_2(z) dz^2.
\]
Then Ricci flat condition has the following solution for the metric functions:
\begin{align*}
(B_{,\,y})^2 &= \Lambda_1 B C_1,\\
(B_{,\,z})^2 &= \Lambda_2 B C_2,\\
f(r) &= 1- \frac{2M}{r} - \frac{1}{4}\left( \Lambda_1+\Lambda_2 \right) r^2.
\end{align*}
More general cases can be easily treated following the same spirit: assume that the bulk 
is $(d+n)$-dimensional, where $d\geq 4$ is the dimension of the black hole slices and $n$ is 
the number of noncompact dimensions. Then the metric should take the form
\[
ds^2 = B(z_{1},...,z_{n})\left( - f(r) dt^2 + \frac{dr^2}{f(r)} + r^2 d\Omega_{d-2}^2 \right) 
+ \sum_{i=1}^{n} C_i (z_i) dz_i^2.
\]
Ricci flatness then imply the following relations:
\begin{align*}
(B_{,\,z_i})^2 &= \Lambda_i B C_i,\\
f(r) &= 1- \frac{2M}{r^{d-3}} - \frac{1}{4}\sum_{i=1}^{n} \Lambda_i  r^2.
\end{align*}
The black hole slices are Tangerlini-Schwarzschild-(A)dS holes in $d$-spacetime dimensions.
Taking the coordinates in which $C_i(z_i)=1$ for all $i$, we have
\[
B(z_{1},...,z_{n}) = \frac{1}{4}\left(c + \sum_i \epsilon_{i} \sqrt{\Lambda_{i}} z_i \right)^2,
\]
where each $\epsilon_{i}$ (for $i=1,...,n$) takes the values $\pm 1$ indepent of the others 
and $c$ is an integration constant. For the metric to be 
real one needs to take all $\Lambda_{i}$ to have equal sign and choose $c$ to be real and purely 
imaginary respectively for positive and negative $\Lambda_{i}$. 

Due to the length of the paper we leave the detailed study on the dynamical and thermodynamical 
properties of these black branes for future works.

\section{Discussions}

In higher spasetime dimensions the family of black hole/brane solutions is significantly enriched. 
In this paper we considered some interesting type of higher dimensional spacetimes which share 
several common properties, in particular the horizons all have non-compact dimensions, and the 
spacetimes considered are all translationally noninvariant along the non-compact dimensions, i.e. 
they are non-uniform. As mentioned in the introduction, study of non-uniform black strings/branes 
is partly triggered by the work of 
Horowitz and Maeda. Though we hoped to find some stable non-uniform solutions, the present yield fails to 
meet this end. However, the method used here can be further adapted to address other interesting cases.
For instance, we list here some of the possible future subjects to study in the future:

\begin{itemize}
\item Adding rotation parameter to the non-uniform black string/ring;

\item Detailed analysis for the stability of each cases, perferable both from the dynamical 
and thermodynamical perspectives;

\item Seeking for non-uniform black ring solutions;

\item Changing the bulk from Ricci flat spacetime to non-Ricci flat ones;

\item Non-uniform black string/ring/branes in supersymmetric models?
\end{itemize}

\bibliography{}

\end{document}